\documentclass[twocolumn,nofootinbib,amsmath,amssymb,preprintnumbers]{revtex4}
\usepackage{graphicx}

\newcommand{\tref}[1]{(\ref{#1})}

\newcommand{\tsemat}[1]{{\mathbf{\textsf{#1}}}}
\newcommand{\Amat}{\tsemat{A}}
\newcommand{\Bmat}{\tsemat{B}}
\newcommand{\Cmat}{\tsemat{C}}
\newcommand{\Dmat}{\tsemat{D}}
\newcommand{\Emat}{\tsemat{E}}
\newcommand{\Tmat}{\tsemat{T}}

\begin{document}

\title{Line Graphs, Link Partitions and Overlapping Communities}
\author{T.S.\ Evans$^{1,2}$}
\author{R.\ Lambiotte$^{1}$\email{r.lambiotte@imperial.ac.uk}}
\affiliation{$^1$ Institute for Mathematical Sciences, Imperial College London, SW7 2PG London, UK \\
$^2$ Theoretical Physics, Imperial College London, SW7 2AZ,
U.K.}

\preprint{Imperial/TP/09/TSE/1, \texttt{arXiv:0903.2181},
Phys.Rev.E \textbf{80} (2009) 016105, DOI:
\texttt{10.1103/PhysRevE.80.016105} }

\date{12th March 2009, revised 3rd June 2009}

\begin{abstract}
In this paper, we use a partition of the links of a network in
order to uncover its community structure. This approach allows
for communities to overlap at nodes, so that nodes may be in
more than one community.  We do this by making a node partition
of the line graph of the original network.  In this way we show
that any algorithm which produces a partition of nodes can be
used to produce a partition of links. We discuss the role of
the degree heterogeneity and propose a weighted version of the
line graph in order to account for this.
\end{abstract}


\maketitle

\section{Introduction}\label{sintro}

Finding hidden patterns or regularities in data sets is a
universal problem which has a long tradition in many
disciplines from computer science \cite{fiedler} to social
sciences \cite{Z77}. For example, when the data set can be
represented as a graph, i.e.\ a set of elements and their
pairwise relationships, one often searches for tightly knit
sets of nodes, usually called communities or modules. The
identification of such communities is particularly crucial for
large network data sets that require new mathematical tools and
computer algorithms for their interpretation. Most community
detection methods find a partition of the set of nodes where
most of the links are concentrated within the communities
\cite{fort,mason}.  Here the communities are the elements of
the partition, and so each node is in one and only one
community.

A popular class of algorithms seek to optimise the modularity
$Q$ of the partition of the nodes of a graph $G$ \cite{N,guimera,Blondel,Rotta,RB06}. The simplest
definition of modularity for an undirected graph, i.e.\ the
adjacency matrix $\Amat$ is symmetric, is \cite{GN}
\begin{eqnarray}
Q(\Amat) = \frac{1}{W}\sum_{C \in \mathcal{P}} \sum_{i,j \in C}
\left[ A_{ij} - \frac{k_i k_j}{W} \right] \,
\label{modAdef}
\end{eqnarray}
where  $W= \sum_{i,j} A_{ij}$ and $k_i = \sum_{j} A_{ij}$ is
the degree of node $i$. The indices $i$ and $j$ run over the
$N$ nodes of the graph $G$.  The index $C$ runs over the
communities of the partition $\mathcal{P}$. Modularity counts
the number of links between all pairs of nodes belonging to the
same community, and compares it to the expected number of such
links for an equivalent random graph in which the degree of all
nodes has been left unchanged. By construction $|Q|\leq 1$ with
larger $Q$ indicating that more links remain within communities
then would be expected in the random model. Uncovering a node
partition which optimises modularity is therefore likely to
produce useful communities.

\begin{figure}
\includegraphics[width=0.22\textwidth]{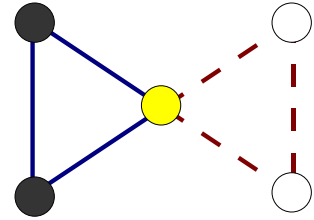}
\caption{(Color online) By partitioning the links of a network into communities,
one may uncover overlapping communities for the nodes by noting
that a node belongs to the communities of its links. In this toy
example, a meaningful partition consists in dividing the links
into two groups (solid blue lines and the dashed red lines). In that case, the
central node belongs to the two communities because it is at the
interface between these link communities.}\label{fbowtiec}
\end{figure}

This node partitioning approach has, however, the drawback that
nodes are attributed to only one community, which may be an
undesirable constraint for networks made of highly overlapping
communities. This would be the case, for instance, for social
networks, where individuals typically belong to different
communities, each characterised by a certain type of relation,
e.g.\ friendship, family, or work.  In scientific collaboration
networks (for example \cite{project2}), authors may belong to
different research groups characterised by different research
interests. Such inter-community individuals are often of great
interest as they broker the flow of information between
otherwise disconnected contacts, thereby connecting people with
different ideas, interests and perspectives
\cite{Burt2004,LP09}.

Only a few alternative approaches have been proposed in order
to uncover overlapping communities of nodes, for example
\cite{PDF,nicosia,fortuna}. Our suggestion is to define
communities as a partition of the links rather than of the set
of nodes. A node may then have links belonging to several
communities and in this it belongs to several communities. The
central node in a Bow Tie graph is a simple example, see Fig.
\ref{fbowtiec}. This link partition approach should be
especially efficient in situations when the nodes of a network
are connected by different types of links, i.e.\ in situations
where the nodes are heterogeneous while the links are very
homogeneous. In the case of the social network mentioned above,
this would occur when the friendship network and work network
of individuals only have a very small overlap.

This paper is organised as follows. In section \ref{sdynmod},
we review a definition of modularity which uses the statistical
properties of a dynamical process taking place on the nodes of
a graph. In section \ref{slinkpart}, we propose three dynamical
processes taking place on the links of the graph and derive
their corresponding modularities, now defined for a partition
of the links of a network. To do so, we make connections to the
concept of a line graph and with the projection of bipartite
networks. In section \ref{sempanal}, we optimise the three
modularities for some examples and interpret our results. In section
\ref{sdiscussion} we conclude and propose ways to improve our
method.

\section{Dynamical formulation of modularity}\label{sdynmod}

To motivate our link partition quality function, let us first
consider how to interpret the usual modularity $Q$
\tref{modAdef} in terms of a random walker moving on the nodes \cite{delvenne,LDB08}.
Suppose that the density of random walkers on node $i$ at step
$n$ is $p_{i;n}$ and the dynamics is given by
\begin{equation}
\label{discrete} p_{i;n+1} = \sum_{j} \frac{A_{ij}}{k_j} \,
p_{j;n} \, .
\end{equation}
From now on, we will only consider networks that are undirected
(the adjacency matrix is symmetric), connected (there exists a
path between all pairs of nodes), non-bipartite (it is not
possible to divide the network into two sets of nodes such that
there is no link between nodes of the same set), and simple
(without self-loops nor multiple links). If the first three
conditions are respected, it is easy to show \cite{Chung} that
the stationary solution of the dynamics is generically given by
$p_i^*=k_i/W$.

Let us now consider a node partition $\mathcal{P}$ of the
network and focus on one community $C \in \mathcal{P}$. If the
system is at equilibrium, it is straightforward to show that
the probability a random walker is in $C$ on two successive
time steps is
\begin{equation}
\label{rw}
 \sum_{i,j \in C}  \frac{A_{ij}}{k_j} \frac{k_j}{W} \, ,
\end{equation}
while the probability of finding two independent walkers at
nodes in $C$ are
\begin{equation}
 \sum_{i,j \in C}  \frac{k_i k_j}{(W)^2} \, .
\end{equation}
This observation allows us to reinterpret $Q$ as a summation
over the communities of the difference of these two
probabilities. This interpretation suggests natural
generalisations of modularity allowing to tune its resolution.
Indeed, $Q$ is based on paths of length one but it can readily
be generalised to paths of arbitrary length as
\begin{eqnarray}
R(\Amat, n) = \frac{1}{W}\sum_{C \in \mathcal{P}} \sum_{i,j \in C}
\left[ (T^n)_{ij} k_j - \frac{k_i k_j}{W} \right] \, ,
\label{stability}
\end{eqnarray}
where $T_{ij} = A_{ij}/k_j$. This quantity is called the stability of the partition \cite{delvenne}. Because $k_j$ is an eigenvector of
 eigenvalue one of $\Tmat$, one can show that the symmetric matrix
$X(n)_{ij}=(T^n)_{ij} k_j$ corresponds to a time-dependent
graph where the degree of node $i$ is always equal to $k_i$.
Therefore $R(\Amat, n)$ can be interpreted as the modularity of
$X(n)_{ij}$, a matrix that connects more and more distant nodes
of the original adjacency matrix $A$ as time $n$ grows \cite{LDB08}. It can
be shown that optimising (\ref{stability}) typically leads to
partitions made of larger and larger communities for increasing
times and that the optimal partition when $n \rightarrow
\infty$ is made of two communities \cite{delvenne,LDB08}.

\begin{figure}
\includegraphics[width=0.33\textwidth]{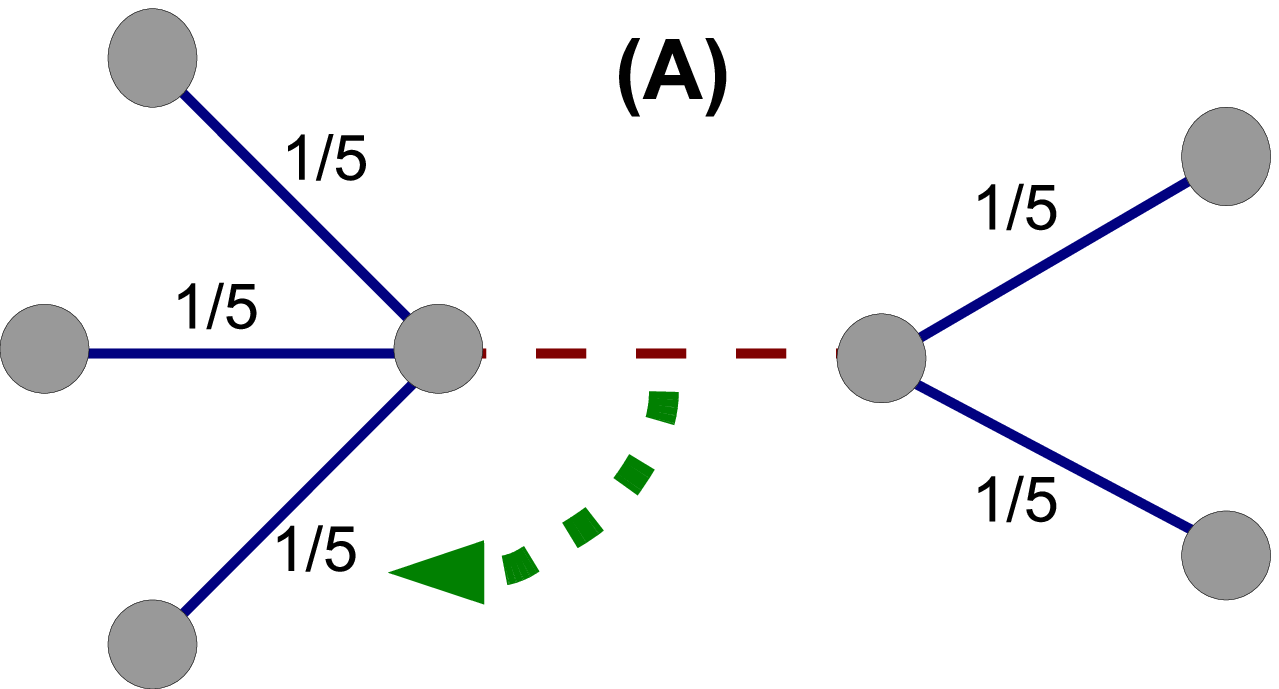}
\hspace{0.05\textwidth}
\includegraphics[width=0.33\textwidth]{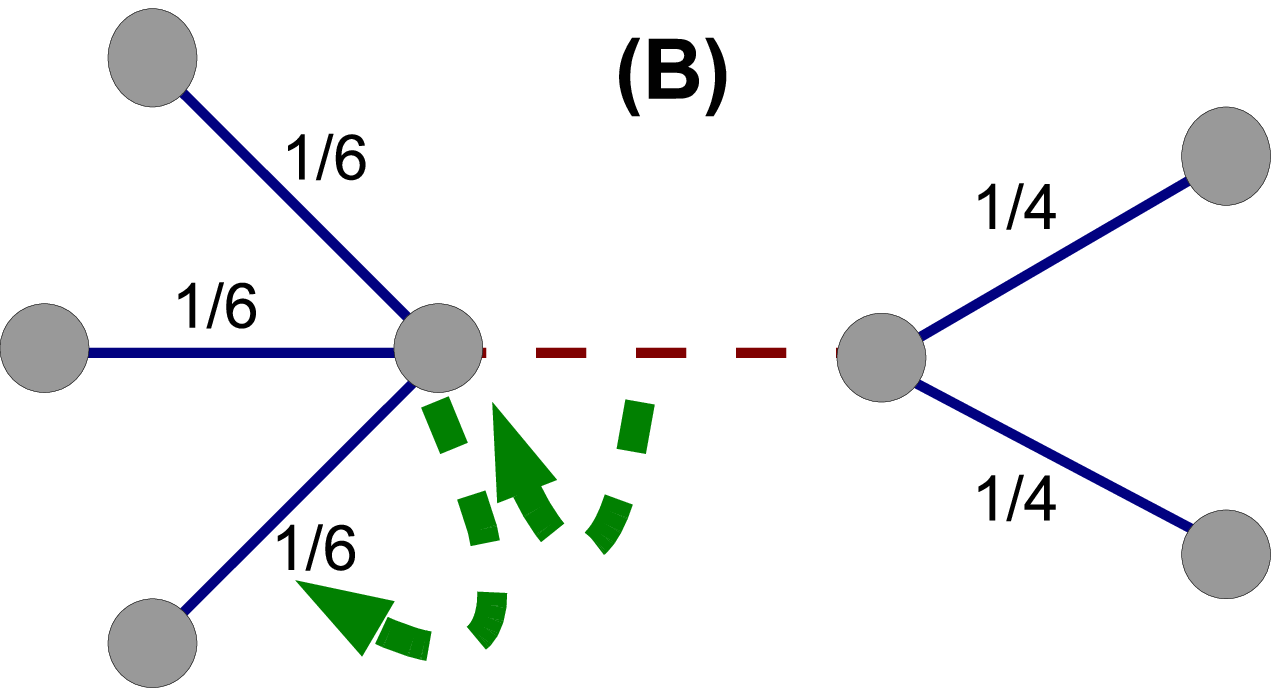}
\caption{(Color online) Illustration of the two types of random walk considered
in this paper. In both cases, the walkers are situated on the
links of a graph, here starting from the central red dashed link.
In (A) the ``Link-Link random walk''  is shown where the walker
jumps (the green dashed arrows) to any of the adjacent links with equal probability. In (B)
a ``Link-Node-Link random walk''  is illustrated.  In this case the walker moves first to
a neighbouring node with equal probability, and then moves on to a new link,
chosen with equal probability from those new links incident at the node.}\label{frandwalks}
\end{figure}

\section{Link partition}\label{slinkpart}

\subsection{Random walking the links}

The above discussion suggests that we should look at a random
walker moving on the links of network in order to define the
quality of a link partition. Such a walker would therefore be
located on the links instead of the nodes at each time $n$ and
move between adjacent links, i.e. links having one node in
common. In the case of the random walk on the nodes
\tref{discrete}, a walker at node $i$ follows one of its links
with probability $1/k_i$, i.e.\ all links are treated equally. However, a
link between nodes $i$ and $j$ is characterised by two
quantities $k_i$ and $k_j$, so a random walk on the links is
more subtle. In the following, we will focus on two different
types of dynamical process that account differently for the
degrees $k_i$ and $k_j$ (see Fig. \ref{frandwalks}).

In the first process, a walker jumps with the same probability
$1/(k_i + k_j - 2)$ to one of the links leaving $i$ and $j$.
When $k_i \neq k_j$, the walker goes with a different
probability through $i$ or $j$, and we therefore call this
process an ``link-link random walk'' (see Fig\
\ref{frandwalks}A).

In the second process, a walker jumps to one of the two nodes
too which it is attached, say $i$, then moves to an link
attached to that node (excluding the link it came from). Thus
it will arrive at an link leaving node $i$ with a probability
$1/( 2 (k_i -1))$, and similarly it will arrive at a link
attached to the other node $j$ with probability $1/( 2 (k_j -
1))$. We will refer to this as a ``link-node-link random walk''
(see Fig\ \ref{frandwalks}B). This process is well-defined
unless the link is a leaf, namely one of its extremities has a
degree one, say $i$. In that case, the walker will jump with a
probability $1/(k_j - 1)$ to one of the links leaving $j$.

These two types of dynamics are different in general except if
the degrees at the extremities $i$ and $j$ of each link are
equal. In the case of a connected graph, this condition is
equivalent to demanding that the graph is regular, i.e. the
degree of all the nodes is a constant. When this condition is
not respected, the link-link random walk favours the passage of
the walker through the extremity having the largest degree. The
difference between the two processes will be maximal when the
network is strongly disassortative, namely when links typically
relate nodes with very different degrees \cite{assortativity}.

\subsection{Projecting the incidence matrix}

\subsubsection{Bipartite structure}

\begin{figure}
\centering
\includegraphics[scale=0.5]{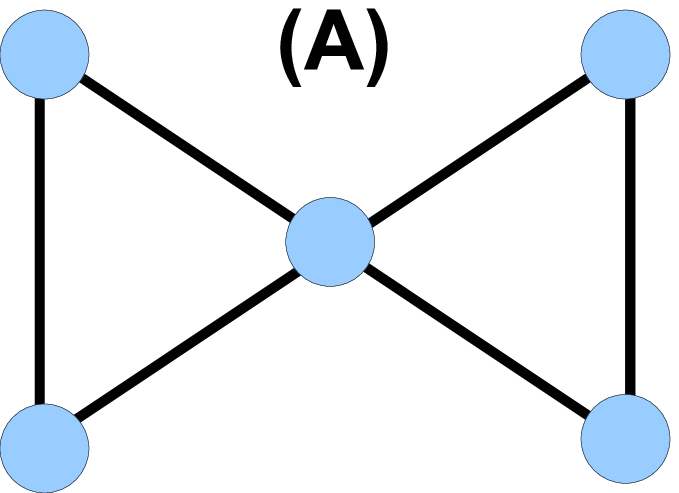}
\hspace{0.05\textwidth}
\includegraphics[scale=0.5]{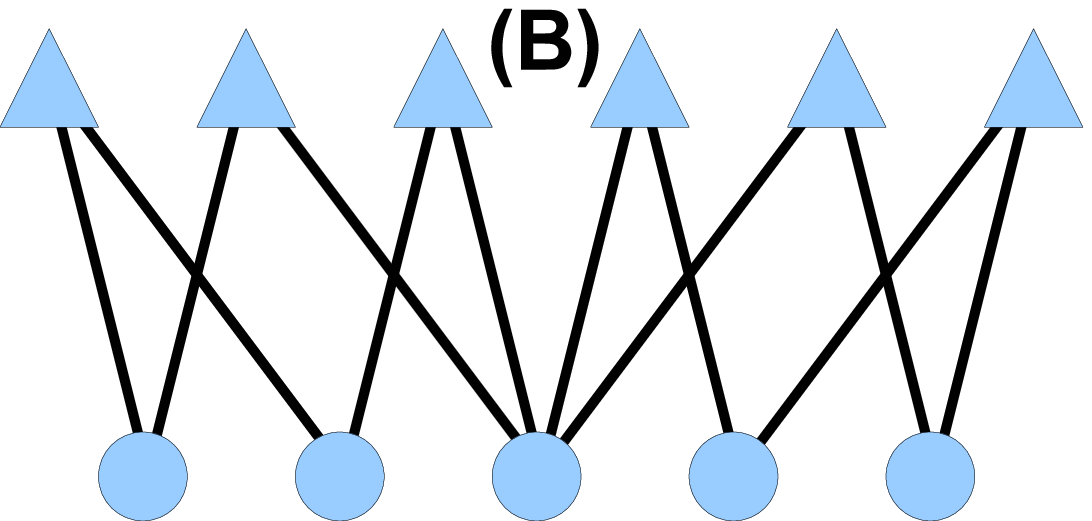}
\\[0.5cm]
\raisebox{0.5cm}{\includegraphics[scale=0.5]{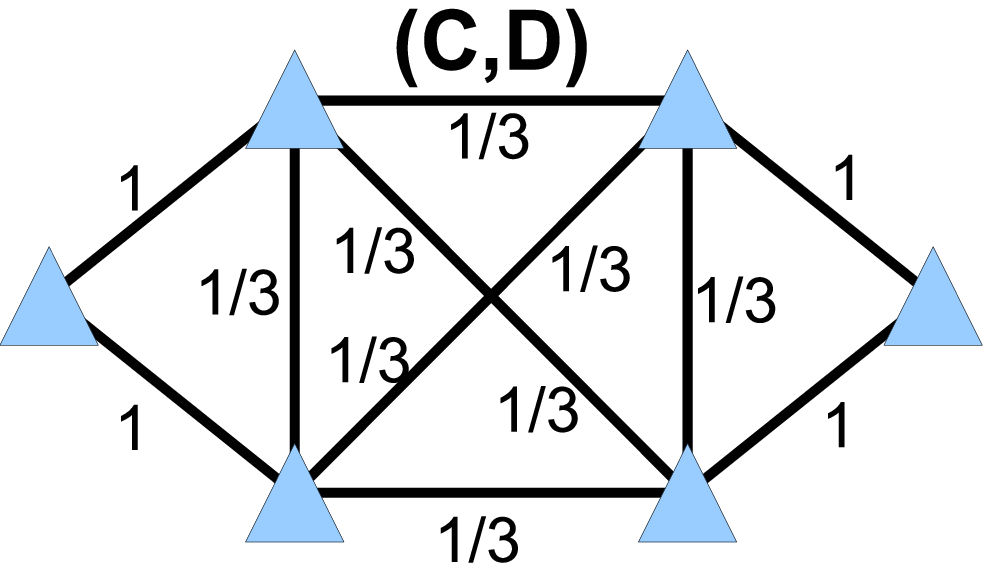}}
\hspace{0.05\textwidth}
\includegraphics[scale=0.5]{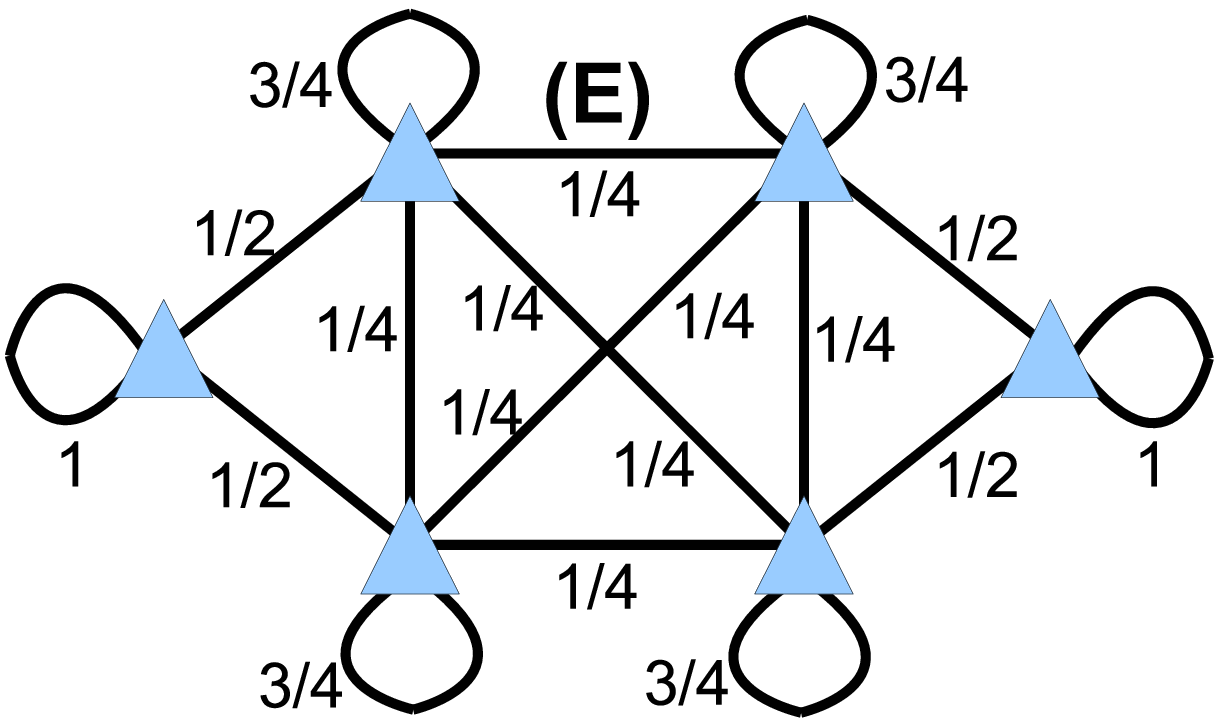}
\caption{(Color online) The information of the Bow Tie graph in (A),
as encoded by the adjacency matrix $\Amat$ of Eqn.\ \tref{adjdef},
has other equivalent graph representations.
In (B) the incidence matrix ($\Bmat$ of Eqn.\ \tref{adjdef}) of the Bow Tie is shown as a
bipartite network, the incidence graph $I(G)$.
The line graph of the Bow Tie, $L(G)$, is the unweighted version of the graph labelled (C,D),
with adjacency matrix $\Cmat$ of Eqn.\ \tref{adjC}.
The weighted version in diagram (C,D) has an adjacency matrix $\Dmat$ of Eqn.\ \tref{adjD}.
The weighted line graph with self loops, labelled ($E$)
has an adjacency matrix $\Emat$ of Eqn.\ \tref{adjDtilde}.
Circles represent entities which correspond to nodes of the original graph,
while triangles come from links in the original graph.
}\label{fbowtieall}
\end{figure}

In order to study these two types of random walk more
carefully, it is useful to represent a network $G$ by its
incidence matrix $\Bmat$. The elements $B_{i \alpha}$ of this
$N \times L$ matrix ($L$ is the number of links) are equal to
$1$ if link $\alpha$ is related to node $i$ and $0$ otherwise.
The incidence matrix of $G$ may be seen as the adjacency matrix
of a bipartite network, $I(G)$ (see Fig.\ref{fbowtieall}B), the
incidence graph\footnote{An incidence graph is usually defined
in terms of the incidence of a set of lines with a set of
points in a Euclidean space of finite dimension. Here we have a
special case where we imbed our graph $G$ in some Euclidean
space of no particular interest, and each link of $G$ is a line
which always intersects with exactly two points.} of $G$ where
the two types of nodes correspond to the nodes and the links of
the original graph $G$. By construction, all the information of
the graph is incorporated in $\Bmat$. For instance, the degree
$k_i$ of a node $i$ and the number of nodes $k_\alpha$ attached
to a link $\alpha$ (always equal to two) are given by
\begin{equation}
 k_i=\sum_\alpha B_{i \alpha} \, , \qquad
 k_\alpha= \sum_i B_{i \alpha}
 \label{kdef}
\end{equation}
The $N \times N$ adjacency matrix $\Amat$ of the graph $G$ can
also be obtained
\begin{equation}
A_{ij}
 = \sum_{\alpha} B_{i \alpha} B_{j \alpha} - k_i \delta_{ij} \, .
\label{adjdef}
\end{equation}
This operation \tref{adjdef} can be interpreted as a projection
of the bipartite incidence graph $I(G)$ onto the unipartite
network $G$ \cite{project1a,project1b}. In a similar way, an adjacency matrix for the
links can be obtained by projecting the bipartite network onto
its links. In the following, we will focus on two standard
types of projection that, as we will show, are directly related
to the two random walks introduced above.

\subsubsection{Line graph}

The simplest way to project a bipartite graph consists of
taking all the nodes of one type for the nodes of the projected
graph. A link is added between two nodes in this projected
graph if these two nodes had at least one node of the other
type in common in the original bipartite graph. The operation
(\ref{adjdef}) is of this type.  When applied to the links
$\alpha$ of the graph $G$, the second type of vertex in the
bipartite incidence graph $I(G)$, it leads to the $L \times L$
adjacency matrix $C$ whose elements are
\begin{equation}
C_{\alpha \beta} = \sum_{i} B_{i \alpha} B_{i \beta} (1-
\delta_{\alpha \beta}). \label{adjC}
\end{equation}
It is easy to verify that this adjacency matrix is symmetric
and that its elements are equal to 1 if two links have one node
in common, and zero otherwise. It is interesting to note that
this adjacency matrix corresponds to another well known graph,
usually called the \emph{line graph} of G \cite{line} and
denoted by $L(G)$ (see Fig.\ref{fbowtieall}C). It is a simple
graph with $L$ nodes. By construction, each node $i$ of degree
$k_i$ of the original graph $G$ corresponds to a $k_i$ fully
connected clique in $L(G)$.  Thus it has $\sum_i k_i(k_i-1)/2 =
O(\langle k^2 \rangle N)$ links. Line graphs have been studied
extensively and among their well-known properties, Whitney's
uniqueness theorem states that the structure of $G$ can be
recovered completely from its line graph $L(G)$, for any graph
other than a triangle or a star network of four nodes
\cite{Whitney}. This result implies that projecting the incidence
matrix onto $L(G)$ does not lead to any loss of information
from the network structure. This is a remarkable result that is
not generally true when projecting generic bipartite networks.

It is now straightforward to express the dynamics of link-link
random walk (Fig.\ref{frandwalks}A) in terms of the projected
adjacency matrix $C$
\begin{equation}
p_{\alpha;n+1} = \sum_{\beta} \frac{C_{\alpha \beta}}{k_\beta}
\, p_{\beta;n}.
\end{equation}
Now $p_{\alpha;n}$ is the density of random walkers on link
$\alpha$ at step $n$, $k_\alpha=\sum_\beta C_{\alpha
\beta}=(k_i + k_j -2)$ and where $i $ and $j$ are the
extremities of $\alpha$. This dynamical process therefore only
depends on the sum of the degrees $i$ and $j$.  The stationary
solution is found to be $p_\alpha^*=k_\alpha/W$, where
$W=\sum_{\alpha \beta} C_{\alpha \beta}$. When $G$ is simple,
then $W= \sum_i (k_i-1)k_i$. By reapplying the steps described
in \cite{LDB08}, it is now straightforward to derive a quality
function for the link partition $\mathcal{P}$ of the graph $G$
\begin{equation}
Q(\Cmat) = \frac{1}{W}\sum_{C \in \mathcal{P}} \sum_{\alpha, \beta \in C}
\left[ C_{\alpha \beta} - \frac{k_\alpha k_\beta}{W} \right].
\label{modCdef}
\end{equation}
 This is just
the usual modularity \tref{modAdef} for a graph with adjacency
matrix $\Cmat$.

As we noted, a single node $i$ in $G$ leads to a connected
clique of $k_i(k_i-1)/2$ links in the line graph $L(G)$. This
seems to suggest that the line graph $L(G)$ gives too much
prominence to the high degree nodes of the original graph $G$.
Our response is to define a weighted line graph whose links are
scaled by a factor of $O(1/k_i)$.

\subsubsection{Weighted line graph}

In order to derive the quality of a link partition associated
to the link-node-link random walk, it is useful to project the
incidence matrix in a different way and to define another graph
$D(G)$ with a symmetric adjacency matrix given by
\begin{equation}
D_{\alpha \beta} = \sum_{i, k_i >1} \frac{B_{i \alpha} B_{i
\beta}}{k_i -1} (1- \delta_{\alpha \beta}). \label{adjD}
\end{equation}
This weighted line graph has the intuitive property that the
degree $k_\alpha=\sum_\beta D_{\alpha \beta}$ of a link
$\alpha$ is equal to two (a link always has two extremities)
unless $\alpha$ is a leaf in $G$ (then $k_\alpha=1$ except for
one trivial case). For example this weighted line graph of the
Bow Tie network is shown in Fig.\ref{fbowtieall}D.  Only if $G$
is regular will this weighted line graph $D(G)$ be equivalent
(up to an overall scale) to the original unweighted line-graph
$L(G)$.

This construction is a well-known method for projecting
bipartite networks. For instance in the case of collaboration
networks \cite{project2} the $(k_i-1)$ normalisation is
justified by the desire that two authors should be less
connected if they wrote a joint paper with many co-authors than
a paper with few authors.

This weighted line graph allows us to write the dynamics of the
link-node-link random walk in a natural way
\begin{equation}
p_{\alpha;n+1} = \sum_{\beta}
\frac{D_{\alpha \beta}}{k_\beta} \, p_{\beta;n}
\end{equation}
 and, by reusing the above arguments to define another quality function
for the link partition $\mathcal{P}$ of a graph
\begin{equation}
Q(\Dmat) = \frac{1}{W}\sum_{C \in \mathcal{P}} \sum_{\alpha, \beta \in C}
\left[ D_{\alpha \beta} - \frac{k_\alpha k_\beta}{W} \right],
\end{equation}
where $W = \sum_{\alpha\beta} D_{\alpha \beta} =
2L-L_\mathrm{leaf}$ is twice the number of links $L$ minus the
number of leaves in the original graph $G$, $L_\mathrm{leaf}$.
Again, this is the same functional form as the usual
modularity, $Q(\Amat)$ of \tref{modAdef}, only the adjacency
matrix has changed.

\subsection{Projection of a node random walk}

The random walks proposed in the previous sections have been
defined on the line graph, and therefore consist of walkers
moving among adjacent links of the original graph $G$. However,
such processes can not be related to the original random walk
(\ref{rw}) on the nodes of $G$, because a walker moving on
links can pass at two subsequent steps through the same node of
$G$ while such self-loops are forbidden in (\ref{rw}). This
observation suggests an alternative approach where the dynamics
would be driven by the original random walk (\ref{rw}) but
would be projected on the links of the network. To do so, let
us assume that a walker has not moved yet and is located at
node $i$. In that case, it is reasonable to assume that all the
neighbouring links of $i$ are connected by a weight $1/k_i$.
The corresponding adjacency matrix $E$ for the links is
therefore given by
\begin{equation}
E_{\alpha \beta} = \sum_{i, k_i >0}
 \frac{B_{i \alpha} B_{i \beta}}{k_i },
 \label{adjDtilde}
\end{equation}
and is based on an unconstrained unbiased two-step random walk
on the bipartite incidence graph $I(G)$\footnote{One might also
try to argue that since an undirected link is both incoming and
outgoing, we might deem it appropriate to allow $\alpha$ to
$\alpha$ transitions in the link-link walk of
Fig.\ref{frandwalks}A. That is we could define an unweighted
line graph with self loops with adjacency matrix
$\tilde{C}_{\alpha \beta} = \sum_i B_{i \alpha} B_{i \beta}$.
Since it differs from the standard unweighted line graph $L(G)$
only by the addition of a self-loop to every node $\alpha$,
this can be interpreted within the scheme of \cite{AFG08} who
add self-loops to control the number and size of communities
found.}. Unlike our previous line graph constructions, $\Cmat$
of \tref{adjC} and $\Dmat$ of \tref{adjD}, this weighted line
graph $E(G)$ has self loops. It is illustrated for the Bow Tie
graph in Fig.\ref{fbowtieall}E. All nodes $\alpha$ in $E(G)$
have strength two, $\sum_\beta E_{\alpha\beta}=2$, reflecting
the fact that the links in the original graph $G$ all have two
ends.

$\Emat$ is constructed when a walker is located on a node and
has not moved yet. The motion of the walker according to
(\ref{rw}) generates a new adjacency matrix, $\Emat_1$, defined
as
\begin{equation}
E_{1;\alpha \beta} = \sum_{i, k_i >0}
 \frac{B_{i \alpha} A_{ij} B_{i \beta}}{k_i k_j},
 \label{adjE1def}
\end{equation}
where we note that $\Emat_1= \Emat\Emat - \Emat$. The
corresponding graph is still regular with $k_\alpha=\sum_\beta
E_{1;\alpha \beta} =2$, and it is again weighted with
self-loops. The quality function associated with this dynamics is simply
\begin{equation}
Q(E_1) = \frac{1}{W}\sum_{C \in \mathcal{P}} \sum_{\alpha, \beta \in C}
\left[ E_{1;\alpha \beta} - \frac{4}{W} \right],
\end{equation}
where again $W=2L$.

This quality function is particularly interesting because it
has a simple relationship to the modularity of the original
graph, $Q(\Amat)$ of \tref{modAdef}.  To show this let us
assign a weight $V_{\alpha c}$ representing the strength of the
membership of link $\alpha$ in community $c$.  Such weights may
be defined and constrained in many ways. For instance, in a
link partition we have $V_{\alpha c} V_{\alpha d} =
\delta_{cd}$ for any $\alpha$, i.e.\ every link $\alpha$
belongs to just one community. In order to translate $V_{\alpha
c}$ into a community structure on the nodes, it is natural to
use the incidence matrix, $\Bmat$ of \tref{adjdef} and to
define the rectangular matrix $V_{i c}$ through
\begin{eqnarray}
 V_{ic} &=& \sum_\alpha \frac{B_{i \alpha}}{k_i} V_{\alpha c} \, .
 \label{ECtoVCdef}
 \end{eqnarray}
If $V_{\alpha c}$ is an link partition then the projected node
community structure $V_{ic}$ is simply the fraction of links in
community $c$ incident at node $i$. Also if $\sum_c V_{\alpha
c}=1$ then so is $\sum_c V_{ic}=1$.

Now using the
definition of the adjacency matrix in \tref{adjdef}, we find
that the modularity of the original graph $G$ for some node
community $V_{ic}$ is
\begin{eqnarray}
Q(E_1; \{V_{\alpha c}\})
 &=&
  \frac{1}{W} \sum_{c,d} \sum_{\alpha, \beta }
  V_{\alpha c}
  \left[ E_{1;\alpha \beta} - \frac{4}{W} \right]
  V_{\beta d}
\\
 &=&
  \frac{1}{W} \sum_{c,d} \sum_{i, j}
  V_{i c}
  \left[ A_{ij} - \frac{k_i k_j}{W} \right]
  V_{j d}
  \\
 &=&
Q(\Amat; \{V_{ic}\})
 \label{ECVCrel}
 \end{eqnarray}
Thus finding modularity optimal link partitions of the line
graph with adjacency matrix $\Emat_1$ of \tref{adjE1def}, is
equivalent to the optimisation of the modularity of the
original graph but with a different constraint on the node
community $V_{ic}$ from that imposed when finding node
partitions.


\section{Empirical analysis}\label{sempanal}

\subsection{Methodology}

In the previous sections, we have proposed three quality
functions $Q(\Cmat)$, $Q(\Dmat)$ and $Q(\Emat_1)$ for the
partition of the links of a network $G$. Each represents a
different dynamical process and therefore explores the
structure of the original graph $G$ in a different way. In
order to tune the resolution of the optimal partitions, it is
straightforward to define the stabilities $R(\Cmat,n)$,
$R(\Dmat,n)$ and $R(\Emat_1,n)$ of the three processes by
generalising the concept of modularity to paths of arbitrary
length (see section II). The optimal partitions of these
quality functions can be found by applying standard modularity
optimisation algorithms to the corresponding line graphs. In
this paper, we have used two different algorithms
\cite{Blondel,Rotta} and have verified that both algorithms
give consistent results.

As a first check, let us look at the Bow Tie graph of Figure
\ref{fbowtiec}. The optimisation of the three quality functions
$Q(\Cmat)$, $Q(\Dmat)$ and $Q(\Emat_1)$ lead to the expected
partition into two triangles, with the values $Q(\Cmat)$=0.1,
$Q(\Dmat)=0.278$, $Q(\Emat_1)=0.167$. In this case, the central
node belongs equally to the two link communities, a situation
which is a far superior way to split the network than a node
partition.  The best node partition gives $Q(\Amat)=0.111$ when
three nodes in one triangle form one community and the
remaining two nodes form a second community.

In order to compare node partitions and link partitions in the
following, we will use the idea of a `boundary link' and a
`boundary node'. A boundary link of a  node partition is one
which connects two nodes from different communities. We will
then define a boundary node of an link partition to be a node
which is connected to links from more than one link community.
Thus the central node of the Bow Tie graph is a boundary node.

\subsection{Karate Club}

\begin{figure*}
\centering
\includegraphics[width=0.3\textwidth]{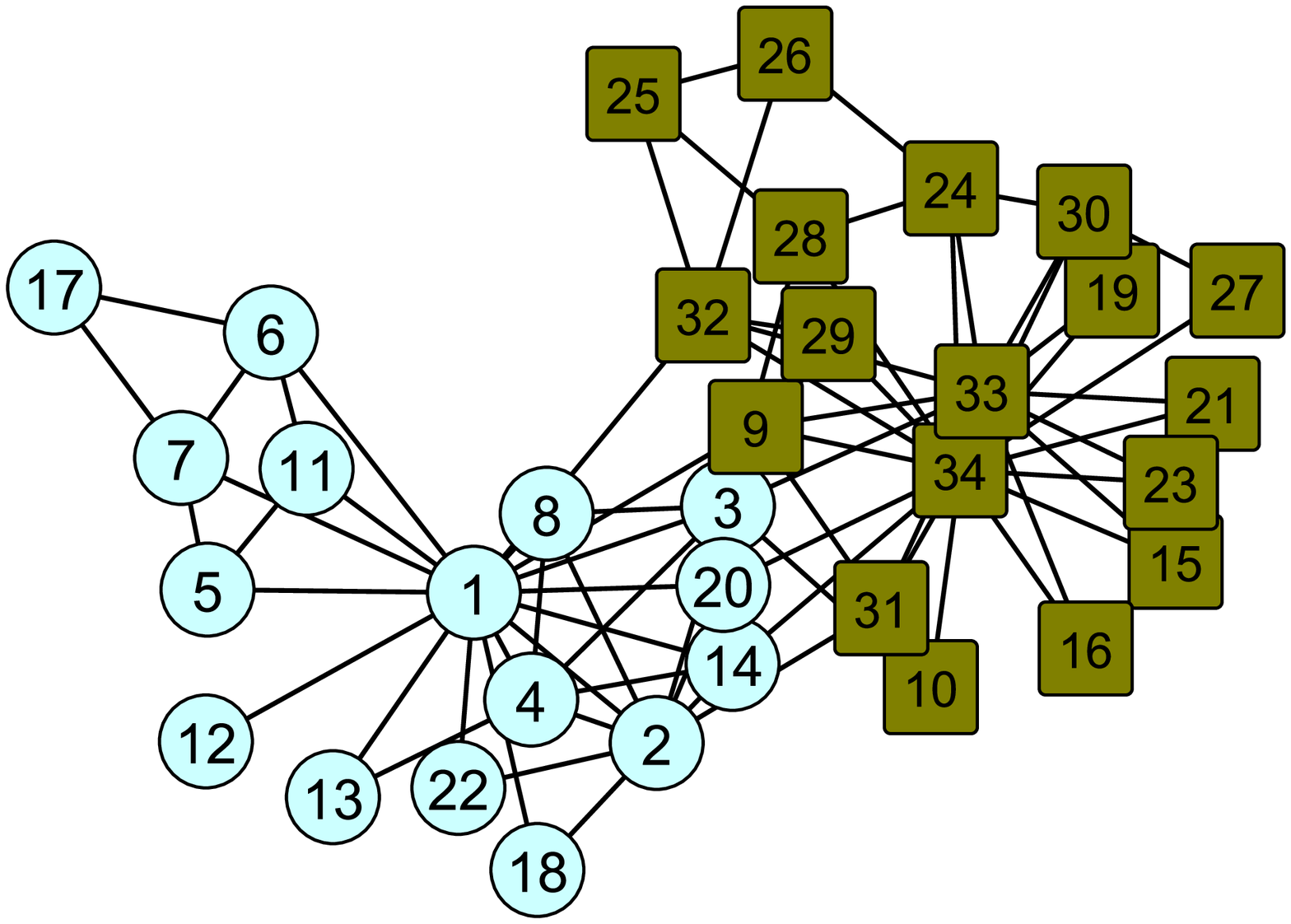}
\includegraphics[width=0.3\textwidth]{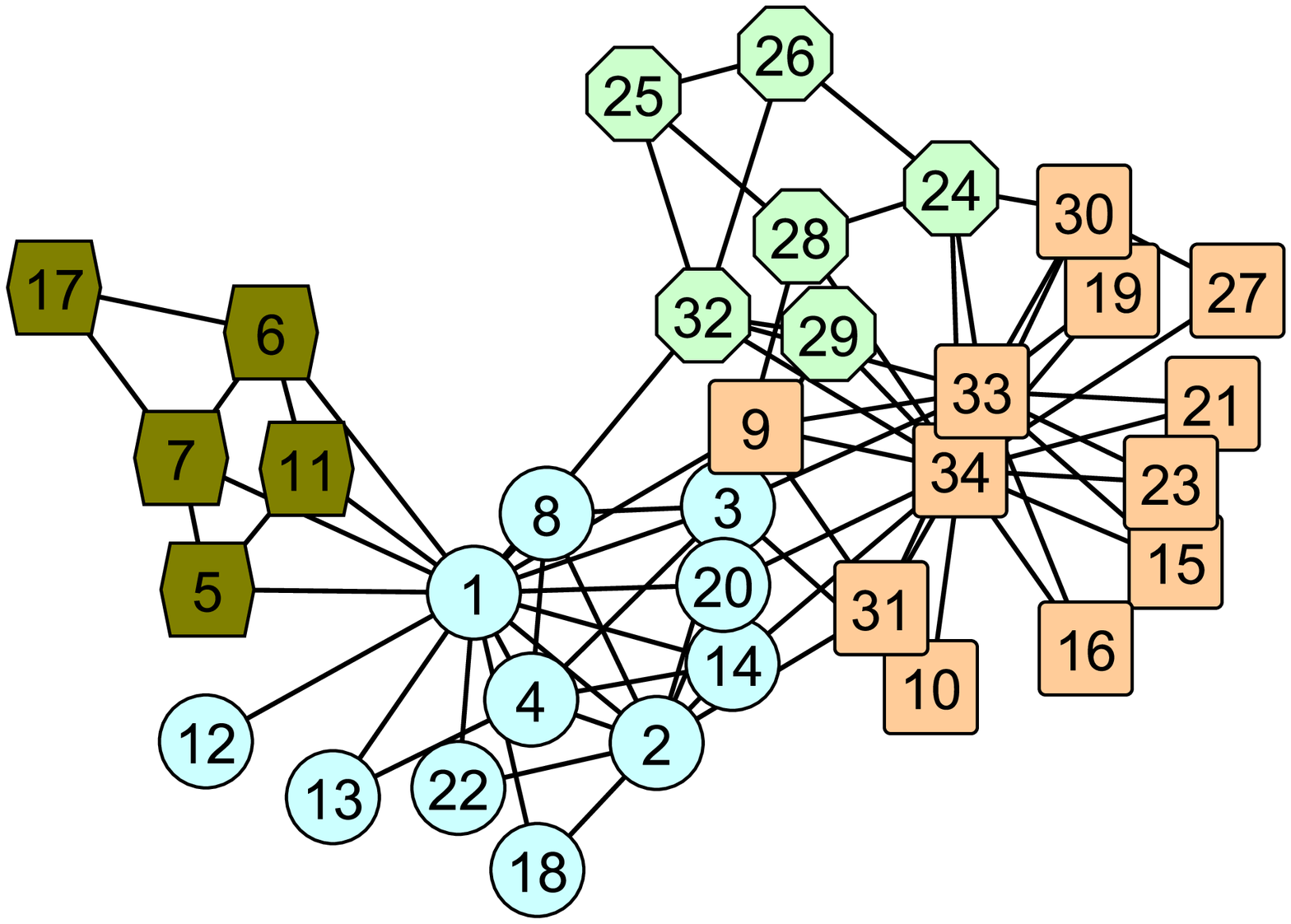}
\caption{(Color online) Optimal node partitions for the
unweighted  Karate Club data of Zachary, notation as in
\cite{Z77}. On the left is the partition into two communities
made by Zachary \cite{Z77} using the Ford-Fulkerson binary
community algorithm \cite{FF56}.  It is also produced by
optimising $R(\Amat,11)$ of \tref{stability}.  The right hand
figure shows the node partition with optimal $Q(\Amat)=0.420$
\cite{AK07} which contains four communities.} \label{fkaratevp}
\end{figure*}

A less contrived graph is the Karate club of Zachary
\cite{Z77}, which is made of thirty four members. Historically
this split into two distinct factions. It is standard to
compare the partition produced by a community detection method
to the actual split of the club. The node partition having the
largest value of modularity $Q(\Amat)=0.420$ contains four
communities, but the resolution can be lowered by optimising
the stability $R(\Amat,n)$ for larger values of $n$. When $n$
is large enough, the optimal partition is always made of two
communities (see Figure \ref{fkaratevp}), e.g.\ $R(\Amat,11)=
0.078$, that agree with Zachary's partition into ``sink'' and
``source'' communities \cite{Z77} using the Ford-Fulkerson
binary community algorithm \cite{FF56}.

\begin{figure*}
\includegraphics[width=0.4\textwidth]{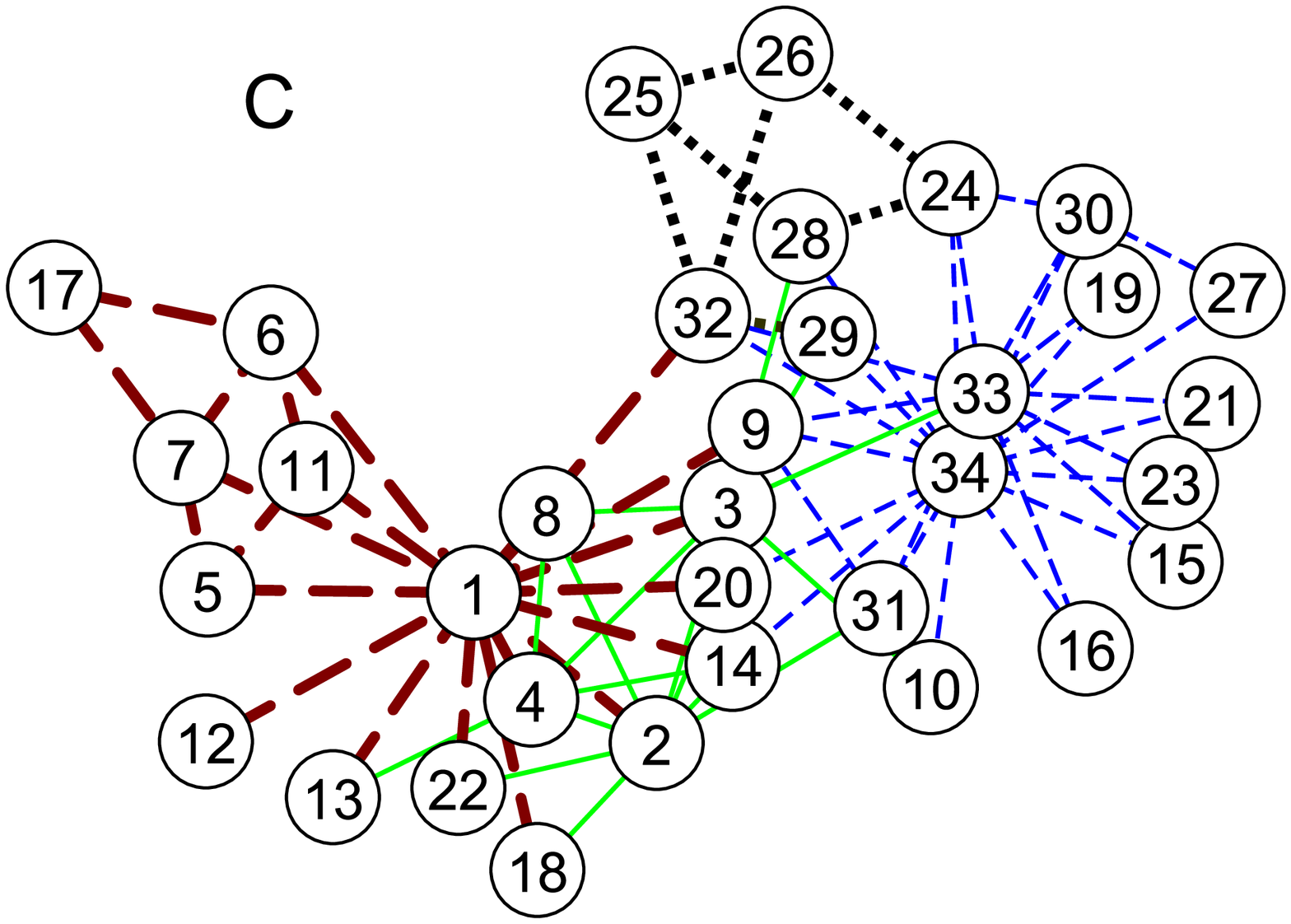} \\
\includegraphics[width=0.4\textwidth]{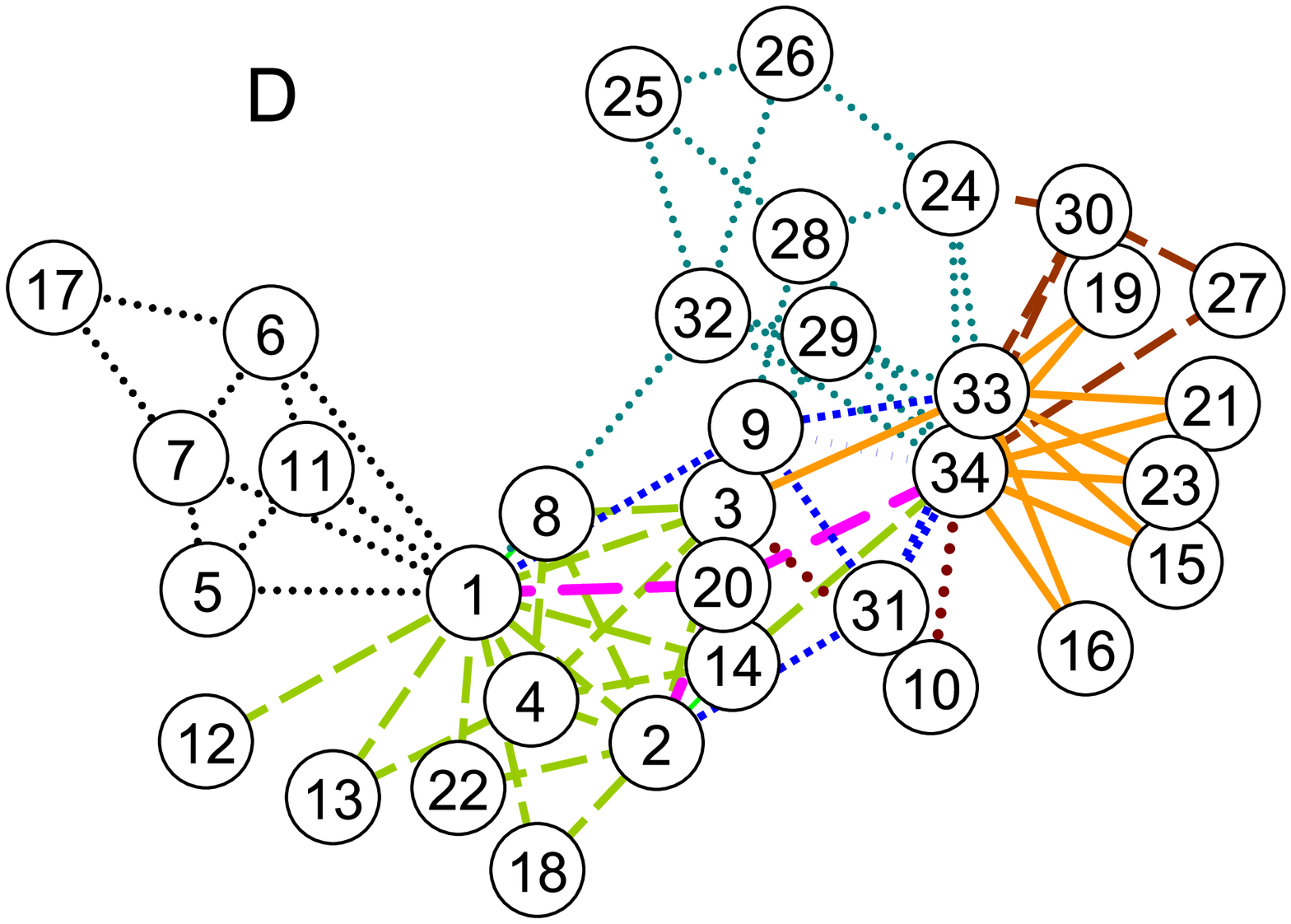} \\
\includegraphics[width=0.4\textwidth]{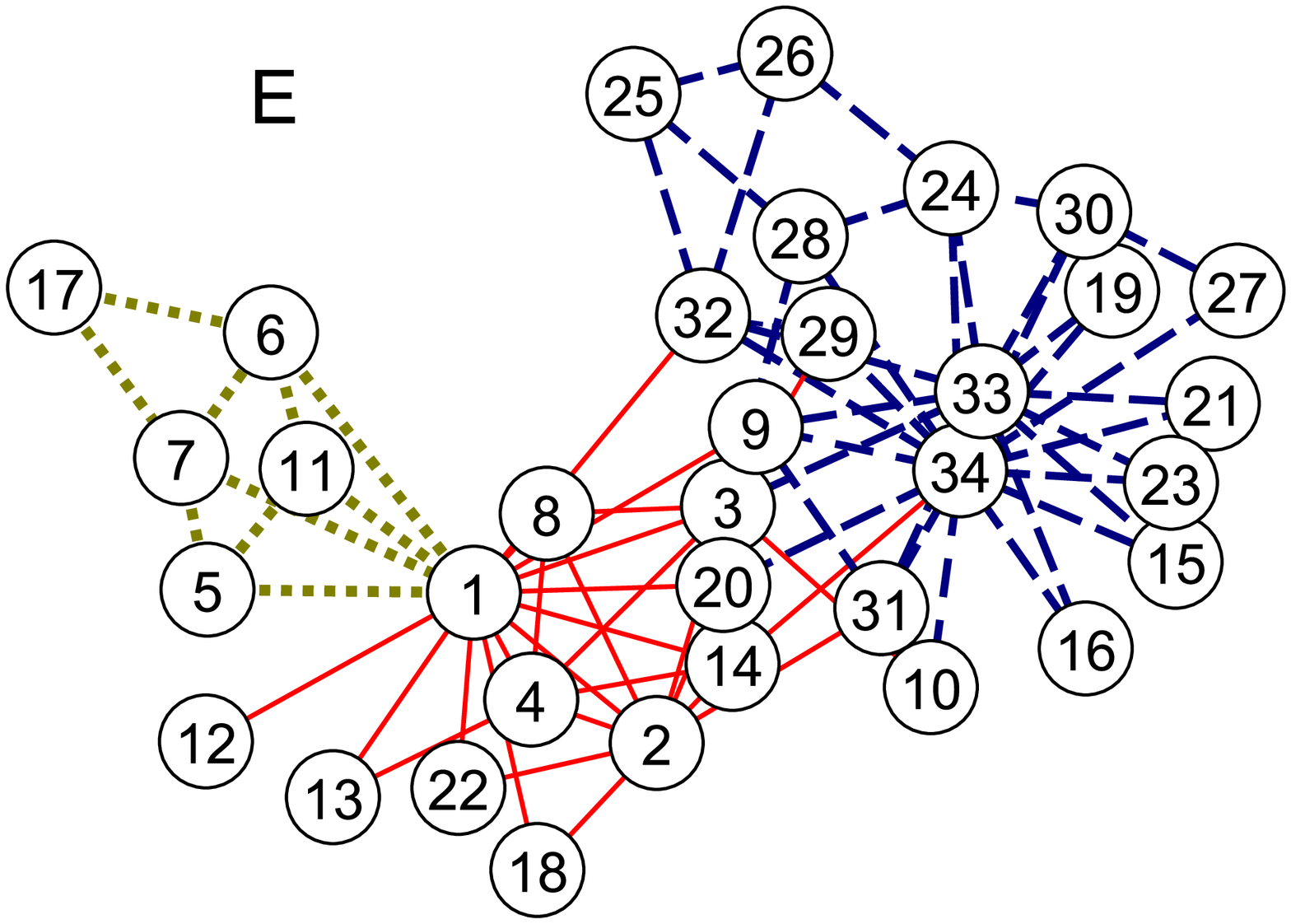}
\caption{(Color online) The optimal link partitions of (C)
$Q(\Cmat)$, (D) $Q(\Dmat)$ and (E) $Q(E_1)$ for the Karate
Club.  They contain $4$, $7$ and $3$ communities respectively.
The two smallest communities in the centre of (D) consist of
the links:
 (a) \{(3,10),  (10,34)\},
 (b) \{(34,20),  (1,20),  (2,20)\}.
}\label{fkarateCDE}
\end{figure*}

The link partitions found by optimising $Q(\Cmat)=0.5$,
$Q(\Dmat)=0.53$ and $Q(\Emat_1)=0.36$ are shown in Fig.
\ref{fkarateCDE}. They are respectively made of $4$, $7$ and
$3$ communities. These three partitions are consistent with the
historical two-way split of the network, as the boundary links
of the two-way partition of Fig.\ \ref{fkaratevp} are always
connected to a boundary node of a link partition. In general,
however, the three optimal partitions are as different as their
corresponding dynamical processes are. The most striking
difference is observed around node $1$. In the node partition
optimising $Q(\Amat)$, this node is connected to several
boundary links and connects the community of nodes
(5,6,7,11,17) to the rest of the network. Such a position is
consistent with the link partitions obtained from $Q(\Dmat)$
and $Q(\Emat_1)$, but not with the link partition optimising
$Q(\Cmat)$. In this latter case, one observes that node 1 is
rather the focus of one of the link communities on the left
hand side in Fig.\ \ref{fkarateCDE}. This difference originates
from the high degree of node 1 which implies that a link-link
random walk is biased to pass through this node (see Fig.\
\ref{frandwalks}), and therefore heavily connects its adjacent
links. This is a general problem of the unweighted line graph
$\Cmat$ that gives too much emphasis to high degree nodes (also
noted in \cite{ABL09}) and therefore tends to produces
communities centred around hubs. Such a problem does not take
place for the weighted line graphs $\Dmat$ and $\Emat_1$, and
in both these cases node 1 is a boundary node, part of several
communities. The main difference between the optimal partitions
of $Q(\Dmat)$ and $Q(\Emat_1)$ is the number of the communities
in each, as expected because the line graph $\Emat_1$ connects
more distance links of the original graph than $\Dmat$. Let us
also note that the optimal partition of $Q(\Emat_1)$ resembles
very much the one of $Q(\Amat)$, as suggested by
\tref{ECVCrel}.

\begin{figure}
\includegraphics[width=0.5\textwidth]{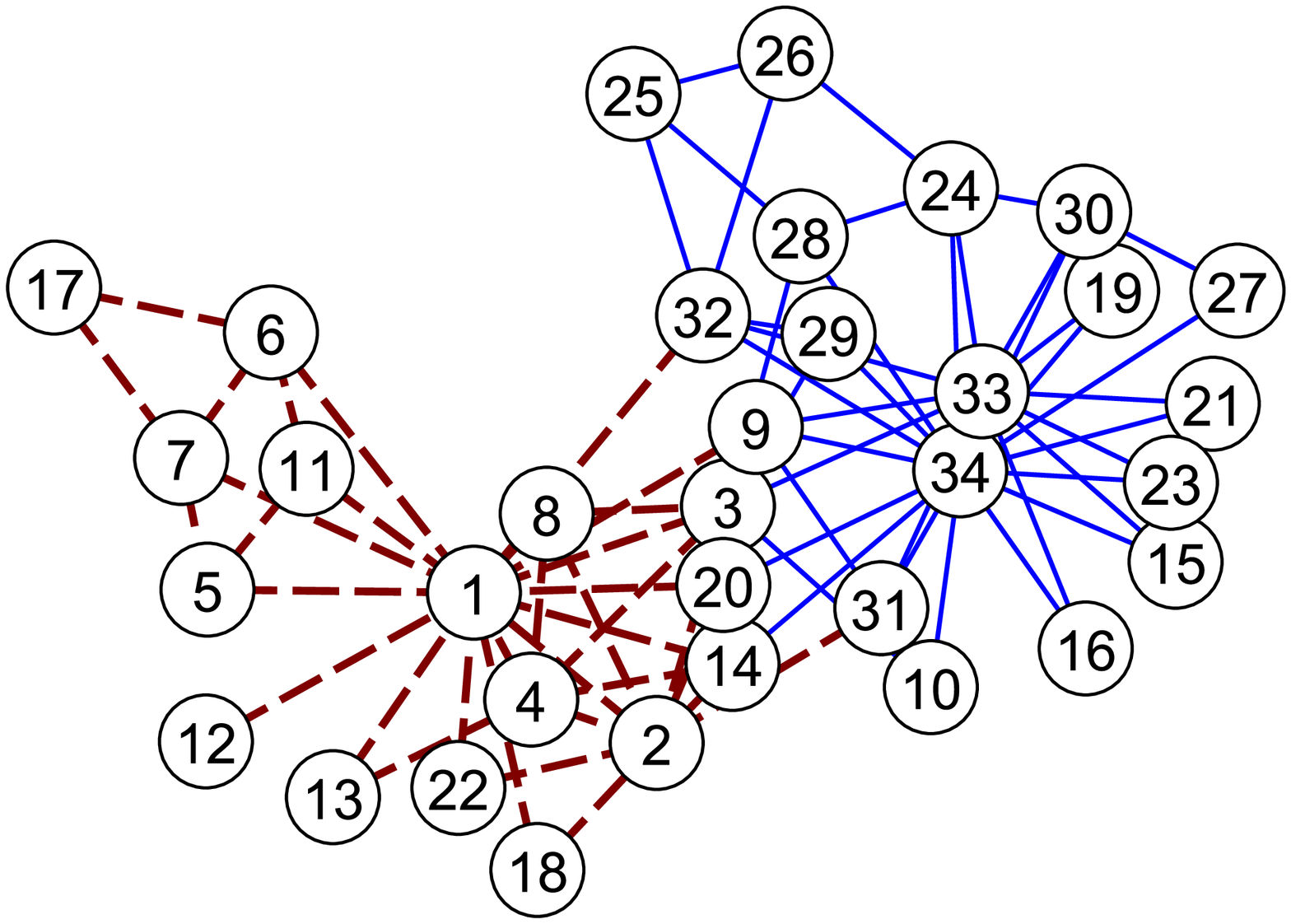}
\caption{(Color online) Optimal link partition into two communities of the stability $R(\Dmat, 10)$
of the Karate club.}\label{fkarateD2}
\end{figure}

Before concluding, let illustrate how longer random walks can
be used to tune the resolution of the link partition. We focus
on the weighted line graph $\Dmat$, whose optimal partition
into seven communities is difficult to compare against the
standard two and four community node partitions of Fig.\
\ref{fkaratevp}. Let us therefore focus on the stability
$R(\Dmat,n)$, which is based on paths of length $n$ of a random
walker on $\Dmat$. As expected, larger and larger communities
are uncovered when $n$ is increased and, when $n$ is large
enough, we obtain a two way link partition (see
Fig.\ref{fkarateD2}) that shows a perfect match with the node
partition shown in Fig.\ref{fkaratevp}.

\subsection{Word Associations}

As a final example, let us use the University of South Florida
Free Association Norms data set \cite{NMS98} to create a simple
network\footnote{We take the sum of the two forward strengths
of all pairs of normed word and add a link only if the total
is greater than $0.025$.  We end up with 5018 words connected
by 58536 links and from this a line graph with 1266910 links is
created.} in the manner of \cite{PDF}.  We obtain a link
partition by optimising the modularity for the weighted line
graph $\Dmat$ of \tref{adjD} but where the null model term
$(k_\alpha k_\beta)/W^2$ has been scaled by a factor of $10.0$
in order to control the resolution \cite{RB06} and in this case
obtain 321 communities in the whole network. The corresponding quality function can be seen as a linear approximation of the stability $R(\Dmat,n)$ \cite{LDB08}. It is easier to optimise for large networks.

\begin{figure}
\centering
\includegraphics[width=0.3\textwidth]{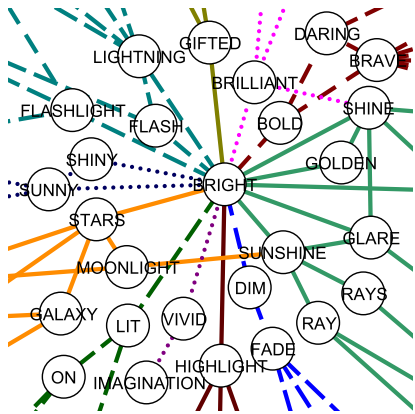}
\caption{(Color online) The simple graph created from the
South Florida Free Association Norms data \cite{NMS98},
in the manner of \cite{PDF}.  The link partition shown
is produced by optimising a modified version of the modularity $Q(\Dmat)$
where the null model factor was $10.0 \times (k_\alpha k_\beta)/W^2$.
This controls the number of communities found \cite{RB06}.
The subgraph shown contains the word `bright' along with nodes
which have at least $90\%$ of their links in one of the communities connected to `bright'.
}\label{fsfwan}
\end{figure}

In Fig.\ref{fsfwan} we show part of the network near the word
`bright' which is part of eleven communities\footnote{The
eleven communities which contain `bright' are well
characterised by the following subsets of words:-
 (`brave', `bold', `daring'),
(`bright', `light', `sunshine'), (`gone', `fade', `dim'),
(`power', `electric', `lightening', `flash'), (`brain',
`intelligence', `brilliant'), (`great', `wonderful', `gifted'),
(`pen', `paper', `highlight'), (`handle', `lit', `on',
`switch', `lever'), (`cloudy', `gray', `shiny', `sunny'),
(`space', `sky', `moonlight', `stars'), (`assume', `illusion',
`imagination', `vivid'). However `bright' has sixteen of its
twenty nine links in the community containing `sunshine' and
`light' with just a single link to eight of its eleven
communities.}. The topology of our communities is much less
constrained than those of k-clique percolation \cite{PDF}
which means we can pick out a wider range of structures. There
are some tight clique-like subsets, e.g.\ the names of the
planets. At the other extreme the method finds more tree like
structures such as the sequence `lit-on-switch-lever-handle'
which is the backbone of another community linked to bright. On
the other hand this flexibility in the structure can produce a
confusing picture since many words are members of several
communities though mostly having just one or two links per
community. For instance for the word `bright', it is linked to
eight of its eleven communities by just one link. However one
can exploit this feature to start to define strength of
membership in different communities.  For instance for
visualisation, we have found it useful to view only those words
which have a large number of links within one community, as in
Fig.\ref{fsfwan}.

\section{Discussion}\label{sdiscussion}

When describing a network, there seems to be a natural tendency
to put the emphasis on its nodes whereas a graph is a both a
set of nodes and a set of links. It is therefore not surprising
that node partitioning has been studied extensively in recent
years while link partitioning has been overlooked so far. In
this paper, we have shown that the quality of a link partition
can be evaluated by the modularity of its corresponding line
graph.  We have highlighted that optimising the modularity of
some of our weighted line graphs uncovers
meaningful link partitions. Our approach has several
advantages. A key criticism of the popular node partitioning
methods is that a node must be in one single community
whereas it is often more appropriate to attribute a node to
several different communities. Link partitioning overcomes this
limitation in a natural way. Moreover, the equivalence of a link
partition of a graph $G$ with the node partitioning of the
corresponding line graph $L(G)$ means that one can use existing
node partitioning code with only the expense of producing a
line graph transformation and an $O(\langle k^2 \rangle/\langle
k \rangle )$ increase in memory to accommodate the larger line
graph. Even the memory cost can be reduced to be $O(1)$ since
we have shown our link partitioning is equivalent to a process
occurring on the links of the original graph $G$, so a line
graph need not be produced explicitly.

Our method can be seen as a generalisation of the popular
k-clique percolation \cite{PDF}, which finds sets of connected
k-cliques.  By way of comparison we find collections of
two-cliques which are more densely connected than expected in
an equivalent null model. Thus the link partitioning of our
paper can be seen as an extension of two-clique percolation
that allows for the uncovering of finer modules, i.e.
two-clique percolation trivially uncovers connected components.
An interesting generalisation would be to apply our approach to
the case of triangles, 4-cliques, etc.  To do so, one has to
replace the incidence matrix (relating nodes and links) by a
more general bipartite graph, representing the membership of
nodes in a clique of interest. Our random walk analysis in
terms of this bipartite graph would then proceed in the same
fashion, and should allow to uncover finer modules than those
obtained by k-clique percolation.

All our expressions also hold for the case of weighted
networks. Even multiedges can be accommodated if we start from
the incidence matrix, $\Bmat$.  However the beauty of our
approach is that any type of graph analysis, be it community
detection or something else, can be applied to a line graph
rather than the original graph.  In this way, one can view a
network from a completely different angle yet use well
established techniques to obtain fresh information about its
structure.

\begin{acknowledgements}
R.L.\ would like to thank M.\ Barahona and V.\ Eguiluz for
interesting discussions, and UK EPSRC for support.  After this
work was finished, we received the paper of Ahn et al.\
\cite{ABL09} who also look at edge partitions but not in terms
of weighted line graphs.
\end{acknowledgements}

\end{document}